\documentclass[twocolumn,a4paper,showpacs,preprintnumbers,amsmath,amssymb,nofootinbib,floatfix]{revtex4}
\def \be {\begin{equation}}

\def \ee {\end{equation}}
\def \bea {\begin{eqnarray}}
\def \eea {\end{eqnarray}}
\usepackage{natbib}
\usepackage{graphicx}
\usepackage{dcolumn}
\usepackage{bm}
\usepackage{epsfig}
\usepackage{xcolor}
\usepackage[normalem]{ulem}

\begin{document}

\title{Constraining a possible time-variation of the speed of light along with the fine-structure constant using strong gravitational lensing and Type Ia supernovae observations}

\author{L. R. Cola\c{c}o$^{1}$} \email{colacolrc@gmail.com}
\author{S. J. Landau$^{2}$} \email{slandau@df.uba.ar}
\author{J. E. Gonzalez$^{3}$} \email{javiergonzalezs@academico.ufs.br}
\author{J. Spinelly$^{4}$} \email{jean.spinelly@hotmail.com}
\author{G. L. F. Santos$^{4}$} \email{gutiery.luis1uepb@gmail.com}

\affiliation{$^1$ Departamento de F\'{i}sica Te\'{o}rica e Experimental, Universidade Federal do Rio Grande do Norte, 59300-000, Natal - RN, Brasil.}
\affiliation{$^2$ Departamento de F\'{i}sica, Facultad de Ciencias Exactas y Naturales, Universidad de Buenos
Aires and IFIBA, Ciudad Universitaria - Pab. I, Buenos Aires 1428, Argentina}
\affiliation{$^{3}$Departamento de Física, Universidade Federal de Sergipe, Aracaju, SE 49100-000, Brasil}
\affiliation{$^4$ Departamento de F\'{i}sica, Universidade Estadual da Para\'{i}ba, 58429-500, Campina Grande - PB, Brasil}

\begin{abstract}

The possible time variation of the fundamental constants of nature has been an active subject of research since the large-number hypothesis was proposed by Dirac. In this paper, we propose a new method to investigate a possible time variation of the speed of light ($c$) along with the fine-structure constant ($\alpha$) using Strong Gravitational Lensing (SGL) and Type Ia Supernovae (SNe Ia) observations. We assume a general approach to describe the mass distribution of lens-type galaxies, the one in favor of the power-law index model (PLAW). We also consider the runaway dilaton model to describe a possible time-variation of $\alpha$. In order to explore the results deeply, we split the SGL sample into five sub-samples according to the lens stellar velocity dispersion and three sub-samples according to lens redshift. The results suggest that it is reasonable to treat the systems separately, but no strong indication of varying $c$ was found.

\end{abstract}
\maketitle

\section{Introduction}

According to standard physics, the fine-structure constant ($\alpha$) that governs the interactions of electrically charged particles is the same throughout the universe in space and time; the speed of light ($c$) in vacuum is the same for all observers; the proportionality constant connecting the gravitational force between two point-like bodies ($G$) is universal; and many others. However, Paul Dirac in 1934 suggested that such constants might not be pure numbers emerging from physical theories, but functions that vary slowly with cosmological time \cite{Dirac:1938mt}. Since then, several theoretical and experimental research allowing space-time variation of fundamental constants has been placed into effect. For instance, some modern alternative theories suggest that the constants of nature might be different in certain places, such as in the extreme gravitational environment around a black hole \cite{hees2020,Antoniou,SilvaHO}. The interest is to know why they have the specific value in what seems to be a ``tuned universe'' (see a complete review in \cite{Uzan:2010pm,Martins:2017yxk}). But one thing is clear, any indication of varying fundamental constants would have deep implications for fundamental physics and cosmology. 

In particular, there have been several proposals to build theories in which the speed of light is dynamical and could have been varying in the past, the so-called Varying Speed of Light (VSL) theories \cite{AlbrechtMagueijo,Barrow,BarrowMagueijo,Magueijo:2003gj}. Furthermore, it was suggested that a modification in the Maxwell-Einstein action could induce light to propagate at speeds higher than the one defined by the metric. However, such a mechanism causes problems with causality and quantum mechanisms \cite{Teyssandier,Adams} (see analyses in \cite{Lee:2020zts,cruz,XuMa,LiuMa,ZhuMa,XuMa2} and references therein). On the other hand, as regards the gravitational sector, many grand-unification theories predict that the gravitational constant $G$ is a slowly varying function of low-mass dynamical scalar fields \cite{Uzan2002,Ballardini,Garcia,Ooba,Mosquera}, while in the electromagnetic sector string-loop effects in string theory models may generate matter couplings for the dilaton (scalar partner of the graviton) that lead to space-time variations of the fine-structure constant ($\alpha \equiv e^2/\hbar c$, where $e$ is the unit electron charge, $\hbar$ is the reduced Planck's constant, and $c$ is the speed of light) \cite{hees,hees2}. In particular, \cite{Damour1,Damour2} developed the runaway dilaton model which assumes the strong coupling limit between matter and the scalar field and is, therefore, able to evade the stringent constraints on violations of the Weak Equivalence Principle (WEP). This model has been used previously to establish constraints of the possible variation in $\alpha$ with galaxy cluster, supernovae type Ia, gravitational lensing, and Sunyaev-Zeldovich data \cite{hol1,hol2,hol3,hol4,hol5}.

As regards the experimental research, authors in Ref. \cite{morel} used matter-wave interferometry to measure the recoil velocity of a rubidium atom that absorbs a photon and determined the value of the fine-structure constant with a relative accuracy of 81 parts per trillion, $\alpha^{-1} = 1/137.035999026(11)$. Nonetheless, several other experiments have been performed throughout the last years with the aim to put stringent constraints on a possible time or spatial variation of $\alpha$. For instance, experiments with atomic clocks provided the tightest constraints on the present variation in $\alpha$ at the level $10^{-17}$ yr$^{-1}$ \cite{2014PhRvL.113u0801G}, while quasar absorption spectra yield $\Delta \alpha/\alpha \propto 10^{-6}$ \cite{Ubachs,Wilczynska,cclee} over a redshift range $0.3 < z < 3.1$.   Besides, constraints on the variation in $\alpha$ in the early universe can be obtained from  cosmic microwave background measurements  \cite{Hart,Aghanim1,Aghanim2,Ade1,Smith2} and primordial nucleosynthesis \cite{2017PhRvC..96d5802M}. Moreover, other limits can be established from white dwarfs \cite{landau1,Bainbridge} and Galaxy clusters \cite{hol1,hol2,hol3,hol4,hol5}; among many others.

As for the Speed of Light (SoL), the value obtained by experiments carried out on the Earth or in our close cosmic surroundings is $c_0 \approx 2.998 \times 10^{5}$ km/s. Precise measurements of $c$ using extragalactic objects are still missing. However, thanks to current technological advances, several observational data enable to measure $c$ at high redshifts accurately. For instance, in Ref.\cite{cao2017a}  $c$ is estimated at the maximum redshift $z=1.70$ using angular diameter distances ($D_A$) from intermediate-luminosity radio quasars calibrated as standard rulers, obtaining $c = 3.039 \pm 0.180 \times 10^5$ km/s (the method was extended by \cite{salzano} afterward). In Ref. \cite{cai2016} a new model-independent method capable of probing the constancy of $c$ throughout a wide redshift range is proposed. The authors argued that deleting the degeneracy between $c$ and the cosmic curvature ($\Omega_k$) makes the test more natural and general. Nevertheless, the method relies on the successful reconstruction of $c$-evolution with redshift and yields $\Delta c/c_0 \sim 1 \%$ at $\sim 1.5 \,\sigma$ confidence level. In Ref. \cite{wang2019}, the authors used a model-independent method to reconstruct the temporal evolution of $c$, and the results were in full agreement with the value measured at $z=0$.

Very recently, the author in Ref. \cite{cao2018} proposed a method that uses the multiple measurements of galactic-scale Strong Gravitational Lensing (SGL) systems with SNe Ia acting as background sources to estimate the speed of light in the distant universe. The results showed $\Delta c/c$ to be at the level $\sim 10^{-3}$. Moreover, \cite{cao2020} also proposed to measure $c$ in the distant universe using multiple different redshift points, but combining SGL Systems and ultra-compact structure observations in radio quasars instead. The results showed precision at the $10^{-4}$  level. Inspired by the previous work, \cite{caonovo} combined the currently available SGL data and the most recent SNe Ia Pantheon sample to perform measurements of $c$ and test its deviation over a wide range of redshift. The advantage of using SNe Ia instead of radio quasars is that there is a large sample available. The results achieved precision at the level $\Delta c/c \sim 10^{-2}$. On the other hand, in Ref. \cite{hol6} a new method to test the invariance of $c$ as a function of redshift combining the measurements of galaxy cluster gas mass fraction, $H(z)$-data from cosmic chronometers, and SNe Ia, is implemented. The analyses indicated a negligible variation of $c$ (see more in \cite{Agrawal} and the references therein).

In this paper, we propose a new method to constrain a possible time variation of $c$ assuming at the same time a possible time variation of the fine-structure constant ($\alpha$). The motivation for assuming this ansatz is very natural since it follows from the dependence of $\alpha$ with $c$. Besides, several theories that predict time variation of the fundamental constants also predict that their variations are related. However, this single ansatz has not been used in the previous works that analyzed a possible variation in $c$ that we mentioned. Besides, we assume the runaway dilaton model to describe both the variations in $\alpha$ and $c$. Our method employs a combination of SGL systems and Type Ia Supernovae (SNe Ia) observations. In particular, we use 111 pairs of observations (SGL - SNe Ia) covering redshift ranges of  $0.075\leq z_l \leq 0.722$ and $0.2551 \leq z_s \leq 2.2649$.

This paper is organized as follows: in section II we discuss the methodology developed to investigate both the invariance of $c$ and a possible time-variation of $\alpha$. In section III we describe the theoretical frameworks. In section IV we present the data set to be used in our analyses, while in section V shows the analyses and discussions. Finally, in session VI, we present the conclusions of this paper.

\section{Methodology}

\subsection{Strong Gravitational Lensing (SGL)}

As it is known, SGL is a purely gravitational phenomenon occurring when the source ($s$), lens ($l$), and the observer ($o$) are at the same line-of-sight to form the so-called Einstein ring, a ring-like structure with angular radius $\theta_E$ \cite{cao2015}.
The two relevant distances, the one from the observer to the lens and the one from the lens to the source, are very large in comparison with the size of the lens galaxy cluster. Therefore, we can assume that the deflection of light occurs in the local Minkowski space-time of the lens, which is perturbed by its gravitational potential \cite{Narayan:1996ba}. This implies that all the physical quantities involved in the light path deviation correspond to their values at the redshift of the lens.

Given the technological advances, SGL systems have been deeply used to investigate many gravitational and cosmological theories. Under the singular isothermal sphere (SIS) model assumption to describe the lens mass distribution, $\theta_E$ is given by \cite{Schneiner,Refsdal}:

\begin{equation}
    \theta_E = \frac{4\pi \sigma_{SIS}^{2}}{c^{2}}\frac{D_{A_{ls}}}{D_{A_s}},
\end{equation}
where $\sigma_{SIS}$ is the velocity dispersion measured under the SIS model assumption, $D_{A_{ls}}$ is the angular diameter distance (ADD) from lens to source, and $D_{A_s}$ is the ADD from observer to source. In this paper, we use $c(z_l)$ instead of just $c$ in order to emphasize that the value of $c$  at the lensing is not equal to its value at our cosmic surroundings ($z=0$). If as a result of the analysis performed in this paper we would get $c(z_l) = c_0$ within statistical and systematic uncertainties, it would confirm either the constancy of $c$ or the standard physics we know on Earth \cite{Cao:2018rzc}. In contrast, if $c(z_l) \neq c_0$, it would be a signal that $c$ is not a fundamental constant, causing further theoretical investigations aiming for an explanation. Thus, we can rewrite Eq. (1) as:

\begin{equation}
    c^2(z_l) = \frac{4\pi \sigma_{SIS}^{2}}{\theta_E}\frac{D_{A_{ls}}}{D_{A_s}}. 
\end{equation}

We assume that the universe can be described by a flat Friedmann-Lemaître-Roberston-Walker metric and we define the comoving distance between the lens and the source as $r_{ls} = r_s - r_l$ \cite{Schneider2}. Moreover, we recall the relations between the comoving distance and the Angular Diameter Distance $D_A$ as follows: $r_s = (1+z_s)D_{A_s}$, $r_l = (1+z_l)D_{A_l}$, and $r_{ls}=(1+z_s)D_{A_{ls}}$. In this way, the following expression can be obtained: 

\begin{equation}
    \frac{D_{A_{ls}}}{D_{A_s}} = 1-\frac{(1+z_l)}{(1+z_s)}\frac{D_{A_l}}{D_{A_s}}.
\end{equation}
Assuming a possible deviation of the Cosmic Duality Distance Relation (CDDR) by $D_{A_i} = D_{L_i}/(1+z_i)^2/\eta(z_i)$, where $\eta(z_i)$ captures any deviation of CDDR, Eq. (3) resumes to:

\begin{equation}
    \frac{D_{A_{ls}}}{D_{A_s}} = 1- \frac{(1+z_s)}{(1+z_l)}\frac{\eta(z_s)}{\eta(z_l)}\frac{D_{L_l}}{D_{L_s}},
\end{equation}
where $D_{L_i}$ is the luminosity distance.  By combining Eq.s (2) and (4), we may obtain:

\begin{equation}
    c^2(z_l) = \frac{4\pi \sigma_{SIS}^{2}}{\theta_E} \left[1-\frac{(1+z_s)}{(1+z_l)}\frac{\eta(z_s)}{\eta(z_l)}\frac{D_{L_l}}{D_{L_s}}    \right].
\end{equation}

\section{Theoretical Framework}

In this paper, we consider that: if $c$ over time is different from its current value ($c_0$), then the fine-structure constant ($\alpha$) over time is different from its current value ($\alpha_0$) as well. We assume that the time variation of $\alpha$ is described by the runaway dilaton model which is a particular case of scalar-tensor theories of gravity.

\subsection{Scalar-Tensor Theory of Gravity}

Let us recall the matter Lagrangian of a scalar-tensor theory of gravity with a non-minimal coupling between the scalar field and matter:

\begin{equation}
    S_{\mathrm{mat.}} = \sum_i \int d^4x \sqrt{-g} h_i(\phi) \mathcal{L}_{i}(g_{\mu \nu}, \Psi_i).
    \label{Sm}
\end{equation}
Here $\mathcal{L}_i$ are the Lagrangians for the different matter fields ($\Psi_i$), and $h(\phi)$ is a function of the extra scalar field. It follows from Eq. (\ref{Sm}) that in this theory the fine-structure constant ($\alpha$) and the CDDR change over cosmological time. Both variations are  unequivocally related as follows:

\begin{eqnarray}
    \frac{\Delta \alpha}{\alpha} && \equiv \frac{\alpha(z)-\alpha_0}{\alpha_0} = \frac{h(\phi_0)}{h(\phi)}-1 = \eta^2(z) - 1. \nonumber \\
    &&\Rightarrow \frac{\alpha(z)}{\alpha_0}=\eta^2(z).
\end{eqnarray}

It is well known that the coupling to the matter fields usually lead to violations of the Weak Equivalence Principle (WEP)  \cite{hees,hees2} which are severely constrained by experimental bounds. For this reason, we choose the runaway dilaton model,  which is able to evade such constraints, to describe the variation in $\alpha$ and $c$. Details are given in the next section.

\subsection{Runaway Dilaton Model}

The so-called runaway dilaton model exploits the string-loop modifications of the (four-dimensional) effective low-energy action.  Unlike other models arising from scalar-tensor theories of gravity, this model is able to evade the stringent experimental constraints of the WEP due to the runaway of the dilaton towards strong coupling. It has been shown \cite{Damour1,Damour2,hees3,Martins2015} that the time variation of $\alpha$ in this model can be expressed as follows:

\begin{eqnarray}
    \frac{\Delta \alpha}{\alpha} && \approx -\frac{1}{40}\beta_{\mathrm{had,0}}\phi_{0}^{'}\ln{(1+z)} \equiv - \gamma \ln{(1+z)} \nonumber \\
    && \Rightarrow \frac{\alpha(z)}{\alpha_0} = 1-\gamma \ln{(1+z)},
\end{eqnarray}
where $\gamma \equiv \frac{1}{40}\beta_{\mathrm{had,0}} \phi_{0}^{'}$, $\beta_{\mathrm{had,0}}$ is the current value of the coupling between the extra scalar field and the hadronic matter\footnote{The relevant parameter of the model is the coupling between the dilaton and hadronic matter.}, and $\phi_{0}^{'} \equiv \frac{\partial \phi}{\partial \ln{a}}$. It is important to emphasize that Eq. (8) can still be considered up to redshift $z \approx 5$ for values of the coupling that saturate the current bounds (see the second panel of Fig. 1 in Ref. \cite{Martins2015}). 

On the other hand, we define a possible time variation of $c$ as $\Delta c/c \equiv [c(z_l)-c_0]/c_0 = c(z_l)/c_0-1$, similarly to $\alpha$. From the fine structure constant definition ( $\alpha \equiv e^2/\hbar c$) and using Eq. (7), we may obtain

\begin{equation}
   \frac{c(z_l)}{c_0} = \frac{e^2}{\hbar \alpha_0 c_0}{\Bigg( \frac{\Delta \alpha}{\alpha} +1 \Bigg)}^{-1}. 
\end{equation}
Next, from  Eqs. (8) and (9) we get an expression for the possible variation of $c$ in this model:

\begin{equation}
    \frac{\Delta c}{c} = \frac{e^2}{\hbar c_0 \alpha_0} \frac{1}{[1-\gamma \ln{(1+z_l)}]} -1.
\end{equation}
We will use this equation to compare the model predictions with SGL systems and SNe Ia data through the method presented here.

\section{Data}

\subsection{Type Ia Supernovae}

We use a sub-sample from Pantheon SNe Ia compilation in order to obtain $D_{L_i}$ for each SGL system. The Pantheon compilation consists of 1048 spectroscopically confirmed SNe Ia covering a wide redshift range of $0.01 \leq z \leq 2.3$ \cite{pantheon}. We construct the $D_{L_i}$ sample from the apparent magnitudes ($m_b$) of Pantheon catalog with $M_b = -19.23 \pm 0.04$ (absolute magnitude) obtained by \cite{R19} and considering the relation
    
    \begin{equation}
    D_{L_i} = 10^{(m_{b_i} - M_{b} - 25)/5} \mathrm{Mpc},
\end{equation}
However, we need the luminosity distances to both lens and source of each SGL system. For that purpose, we carefully select SNe Ia with redshifts obeying the criteria $|z_l - z_{\rm SNe Ia}| \leq 0.005$ and $| z_s - z_{\rm SNe Ia}| \leq 0.005$. Hence, we calculate the weighted average with the corresponding error by:

\begin{equation}
\bar{D}_L = \frac{\sum_i {D_L}_i/\sigma_{{D_L}_i}^{2}}{\sum_i 1/\sigma_{{D_L}_i}^{2}},
\end{equation}

\noindent and 
\begin{equation}
\sigma_{\bar{D}_{L}}^2 = \frac{1}{\sum_i 1/ \sigma_{{D_L}_i}^{2}}.
\end{equation}
    
\noindent where $\sigma_{m_{b_i}}^2$ and $\sigma_{D_{L_i}}^{2} = (\partial D_{L_i}/ \partial m_{b_i})^2\sigma_{m_{b_i}}^2$ are the apparent magnitude and luminous distance errors, respectively (see Fig. 1). 
\begin{figure}[htb!]
\centering	
\includegraphics[scale=0.55]{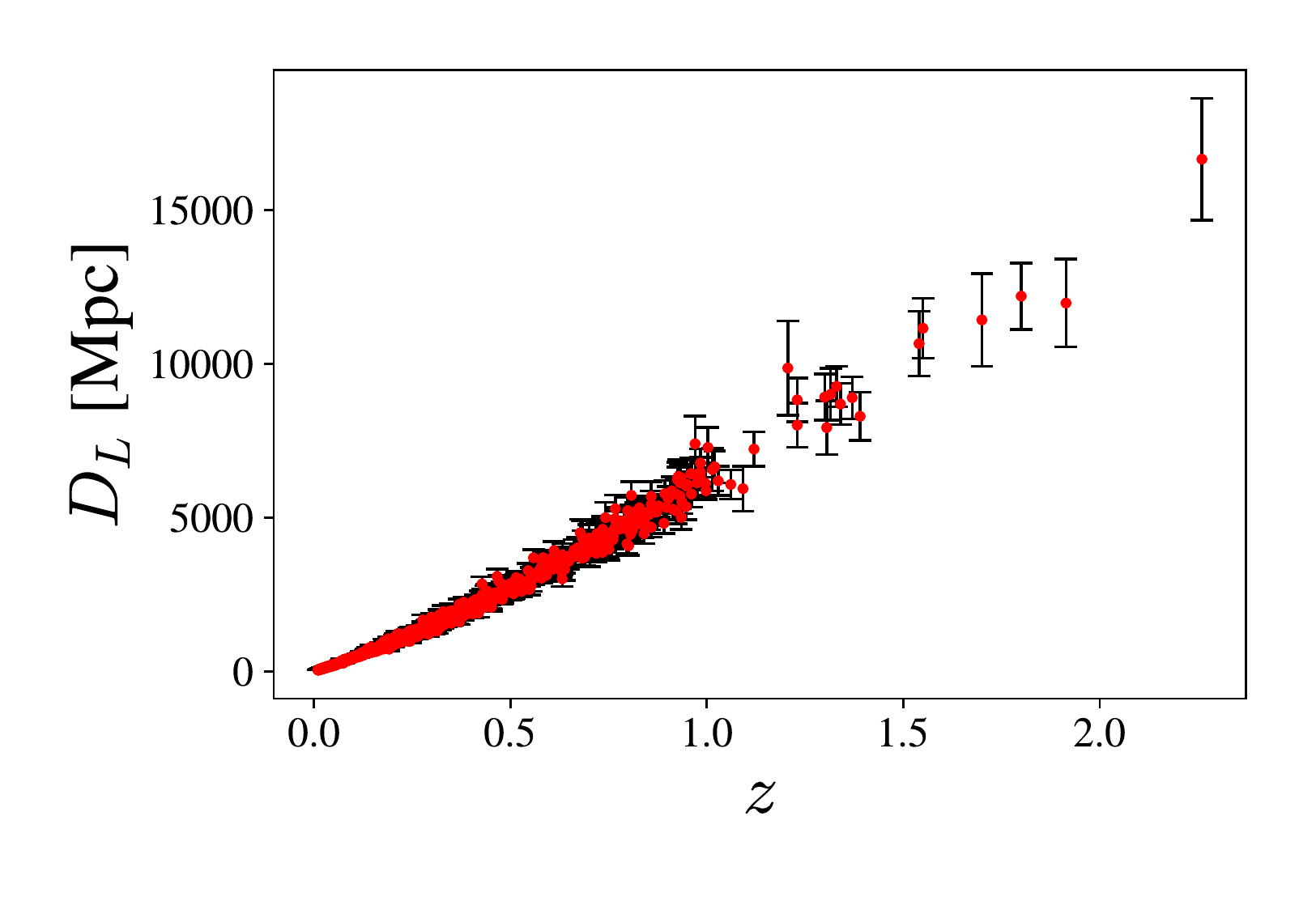}
\caption{Luminosity distances of spectroscopically-confirmed SNe Ia from Pantheon compilation. We constructed $D_{L_i}$ sample from the apparent magnitudes ($m_b$) and $M_b = -19.23 \pm 0.04$ obtained by \cite{R19}.}
\label{dlriess}
\end{figure}

\subsection{Strong Gravitational Lensing Systems}

We consider a specific catalog containing 158 confirmed sources of SGL compiled by \cite{leaf2018}. Such a compilation contains 118 SGL systems identical to the compilation of \cite{cao2015} derived from the SLOAN Lens ACS \cite{bolton,augerMW}, BOSS Emission-line Lens Survey (BELLS) \cite{Brownstein}, and Strong Legacy Survey SL2S \cite{RuffAJ,ASonnenfeld,ARGavazzi,TSonnenfeld}. We note that the lens galaxies from such a catalog should guarantee the validity of the SIS hypothesis. Such validity is secured by the selection of early-type lens galaxies, and those should not have evident substructures or close massive companions (either physical or projected ones) \cite{cao2015,caonovo}. The catalog also contains 40 new systems recently discovered by SLACS and pre-selected by \cite{Yshu} (see Table I in Ref. \cite{leaf2018}).  

Following recent analyses of the lens mass distribution models, we consider the so-called power-law (PLAW) model. It essentially assumes a spherically symmetric mass distribution with more general power-law index $\Upsilon$, specifically $\rho \propto r^{-\Upsilon}$ ($\rho$ is the total mass distribution and $r$ is the spherical radius from the center of the lensing galaxy.). This procedure occurs because several recent studies have shown that the slopes of density profiles of individual galaxies exhibit a non-negligible deviation from SIS model \cite{Koopmans,Auger2010,Barnabe2011,Sonnenfeld2013}. Therefore, assuming that the velocity anisotropy can be ignored and solving the spherical Jeans equation, we can rescale the dynamical mass inside the aperture of size $\theta_{ap}$ projected to the lens plane and obtain:

\begin{equation}
    \theta_E = \frac{4\pi \sigma_{ap}^{2}}{c^2}\frac{D_{A_{ls}}}{D_{A_s}} \Bigg(    \frac{\theta_{E}}{\theta_{ap}} \Bigg)^{2-\Upsilon} f(\Upsilon),
\end{equation}
where $\sigma_{ap}$ is the stellar velocity dispersion inside the aperture $\theta_{ap}$ and

\begin{eqnarray}
    f(\Upsilon) &=& -\frac{1}{\sqrt{\pi}} \frac{(5-2\Upsilon)(1-\Upsilon)}{3-\Upsilon}\frac{\Gamma (\Upsilon-1)}{\Gamma (\Upsilon -3/2)} \nonumber \\
     && \times \left[     \frac{\Gamma (\Upsilon/2 - 1/2)}{\Gamma (\Upsilon/2)}\right]^2.
\end{eqnarray}
If $\Upsilon=2$ we recover the SIS model. By combining Eq.s (5), (7), (8), and (14) we  obtain:

\begin{eqnarray}
   && c^2(z_l) = \frac{4\pi \sigma_{ap}^{2}}{\theta_E}f(\Upsilon) \Bigg( \frac{\theta_{E}}{\theta_{ap}}   \Bigg)^{2-\Upsilon} \nonumber \\
    && \times \left[ 1-\frac{(1+z_s)}{(1+z_l)} \frac{D_{L_l}}{D_{L_s}}\sqrt{\frac{\alpha(z_s)}{\alpha(z_l)}}  \right].
\end{eqnarray}

In order to test the invariance of $c$ using SGL systems and SNe Ia, an important issue needs clarification: the central velocity dispersion of the lens $\sigma_{ap}$ depends on the value of $c$. The most ordinary approach to determine velocity dispersions is to compare a galaxy spectrum with a star spectrum taken through the same spectrograph with the same adjustments \cite{rixwhite,cappellari}. In this context, the spectral lines become wider due to Doppler Effect, and $\sigma_{ap}$ might be inferred by the observed quantity $c_0\Delta \lambda/\lambda$. Only dimensionless quantities might have invariant meaning from a theoretical physics context. Thus, introducing the dimensionless quantity $\Delta c/c (z_l)$, we obtain:

\begin{eqnarray}
 &&\frac{\Delta c}{c} (z_l) \equiv \frac{c(z_l)-c_0}{c_0} =\frac{\sigma_{ap}}{c_0}\sqrt{\frac{4\pi}{\theta_E}K} -1,
\end{eqnarray}
where
\begin{equation}
  K \equiv f(\Upsilon)  \Bigg(  \frac{\theta_E}{\theta_{ap}}  \Bigg)^{2-\Upsilon}\left[   1-\frac{(1+z_s)}{(1+z_l)}\frac{D_{L_l}}{D_{L_s}}\sqrt{\frac{\alpha(z_s)}{\alpha(z_l)}}  \right].
\end{equation}
Here $\Delta c/c (z_l)$ captures the deviation of $c(z_l)$ from $c_0$. We should also point out that recently some authors have performed analyses considering  a possible time evolution of the mass density power-law index \cite{hol4,cao2018,Koopmans,Auger2010,Amante2019xao,ChenShu,CaoBiesiada2,Holanda9}. The results indicated that : {\bf i)} no strong evolution of $\Upsilon$ has been found; {\bf ii)} it is essential to use low, intermediate, and high-mass galaxies separately in any cosmological analyses. Therefore, we consider $\Upsilon$ as a free parameter in this paper. Most of the relevant information necessary to obtain Eq. (17) can be found in Table 1 of Ref. \cite{leaf2018}.

As mentioned previously, our SGL sample consists of 158  points covering a wide redshift range. However, not all systems have the corresponding pair of luminosity distances via SNe Ia obeying the criteria. For this reason, we excluded those systems plus the system J0850-03471\footnote{It deviates by more than $5\sigma$ from all the considered models \cite{leaf2018}}. Therefore, we finish with 111 pairs of observations {covering redshift ranges of $0.0625 \leq z_l \leq 0.722$ and $0.2172 \leq z_s \leq 1.550$.} 

\section{Analysis and Discussions}

We use Markov Chain Monte Carlo (MCMC) methods to estimate the posterior probability distribution functions (pdf) of free parameters supported by {\bf emcee} MCMC sampler \cite{Foreman}. To perform the plots, we used the GetDist Python package. The likelihood is given by:

\begin{equation}
    \mathcal{L} (\mathrm{Data}|\Vec{\Theta}) = \prod \frac{1}{\sqrt{2\pi}\sigma_{\mu}} e^{ -\frac{1}{2}\chi^2}, 
\end{equation}
where

\begin{equation}
    \chi^2 = \sum_i \frac{[\Delta c_i/c_i (z_l) - \Delta c/c]^2}{\sigma_{\Delta c_i/c_i(z_l)}^{2}},
\end{equation}
$\Delta c/c$ and $\Delta c_i/c_i (z_l)$ are given, respectively, by Eq.s (10) and (17), and
\begin{equation}
    \sigma_{\Delta c_i/c_i(z_s)}^{2} = \sigma_{\theta_{E_i}}^{2} + \sigma_{\sigma_{ap_i}}^{2} + \sigma_{D_{L_{l_i}}}^{2} + \sigma_{D_{L_{s_i}}}^{2}
\end{equation}
the associated error. Following the approach taken in Ref. \cite{Grillo}, Einstein’s radius uncertainties follow $\sigma_{\theta_E} = 0.05 \theta_E$ ($5\%$ for all systems). Moreover, we replace $\sigma_{ap}$ by $\sigma_0$ in Eq. (17). This procedure makes the ratio $D_{A_{ls}/D_{A_s}}$ more homogeneous for a sample of lenses located at different redshifts \cite{cao2015}.

The pdf is proportional to the product between likelihood and prior ($P(\Vec{\Theta})$), that is,

\begin{equation}
    P(\Vec{\Theta}|\mathrm{Data}) \propto \mathcal{L}(\mathrm{Data}|\Vec{\Theta})\times P(\Vec{\Theta}).
\end{equation}
In our analyses, we assume flat prior for the free parameters ($\Vec{\Theta} = (\gamma, \Upsilon$)).

We obtain: $\gamma=-0.44 \pm 0.105$ and $\Upsilon = 1.88 \pm 0.075$ with $\chi_{\mathrm{red}}^{2} \approx 2.379$ at $1\sigma$ of confidence level for the whole sample. As the random variation in galaxy morphology is almost Gaussian, the authors of Ref. \cite{leaf2018} found that it is necessary to add an intrinsic error $\sigma_{\mathrm{int}} \approx 12.22\%$ to have $68.3\%$ of the observations lying within $1\sigma$ of the best-fit $\omega$CDM model. Therefore, adding this intrinsic error, our results for the whole sample are  $\gamma = -0.21 \pm 0.295$ and $\Upsilon = 1.81 \pm 105$ with $\chi_{\mathrm{red}}^{2} \approx 0.692$ ($1\sigma$) (see Fig. 2).

\begin{figure}[htb!]
\centering	
\includegraphics[scale=0.8]{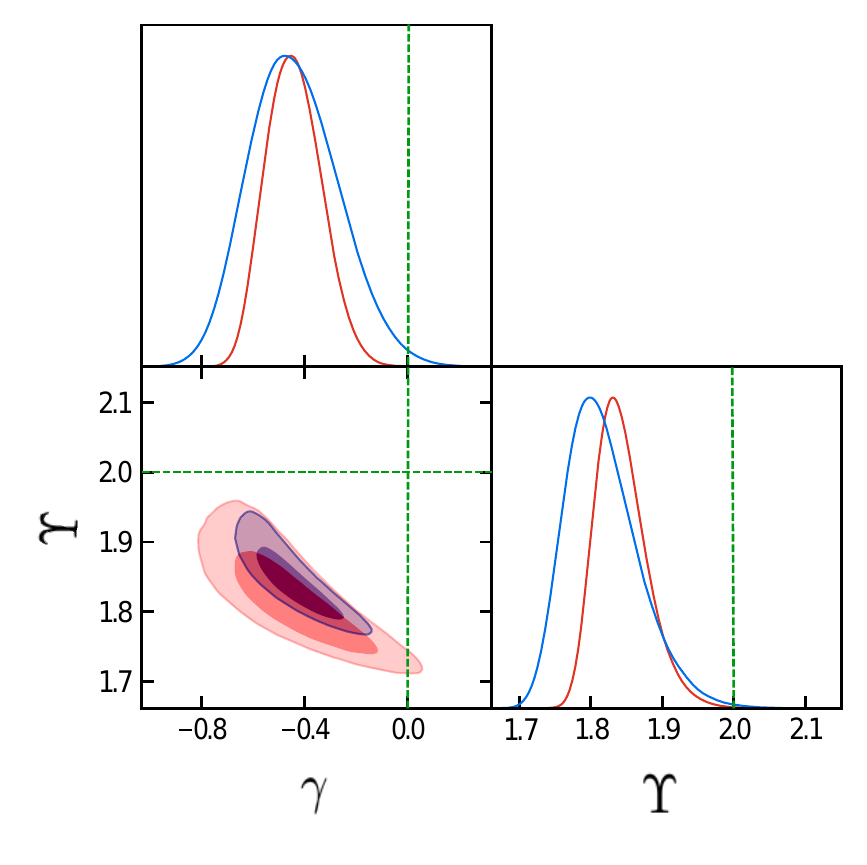}
\caption{Posterior probability distribution functions of $\gamma$ and $\Upsilon$ with (red outline) and without (blue outline) intrinsic error considering $111$ pairs of SGL-SNe Ia. The green vertical dashed lines correspond to the limits $\gamma = 0$ and $\Upsilon = 2.0$.}
\label{dlriess}
\end{figure}

In order to check for the consistency of our results, we split the sample into five sub-samples according to the lens stellar velocity dispersion and three sub-samples according to the lens redshift (S$i$). In this way we obtain: 28 systems with $ \sigma_0 <200 $ {\rm km/s} ({\bf S1}), 40 systems with $200 \leq \sigma_0 \leq 250 $ {\rm km/s} ({\bf S2}), 31 systems with $250 < \sigma_0 < 300 $ {\rm km/s} ({\bf S3}), 11 systems with $\sigma_0 \geq 300$ {\rm km/s} ({\bf S4}), 72 systems with $200 \leq \sigma_0 < 300$ {\rm km/s} ({\bf S5}), 55 systems with $z_l <0.2$ ({\bf S6}), 17 systems with $z_l >0.4$ ({\bf S7}), and 39 systems with $0.2 \leq z_l \leq 0.4$ ({\bf S8}). The results are summarized in Table 1 and shown in Figs. 3 and 4).

\begin{figure}[htb!]
\centering	
\includegraphics[scale=0.9]{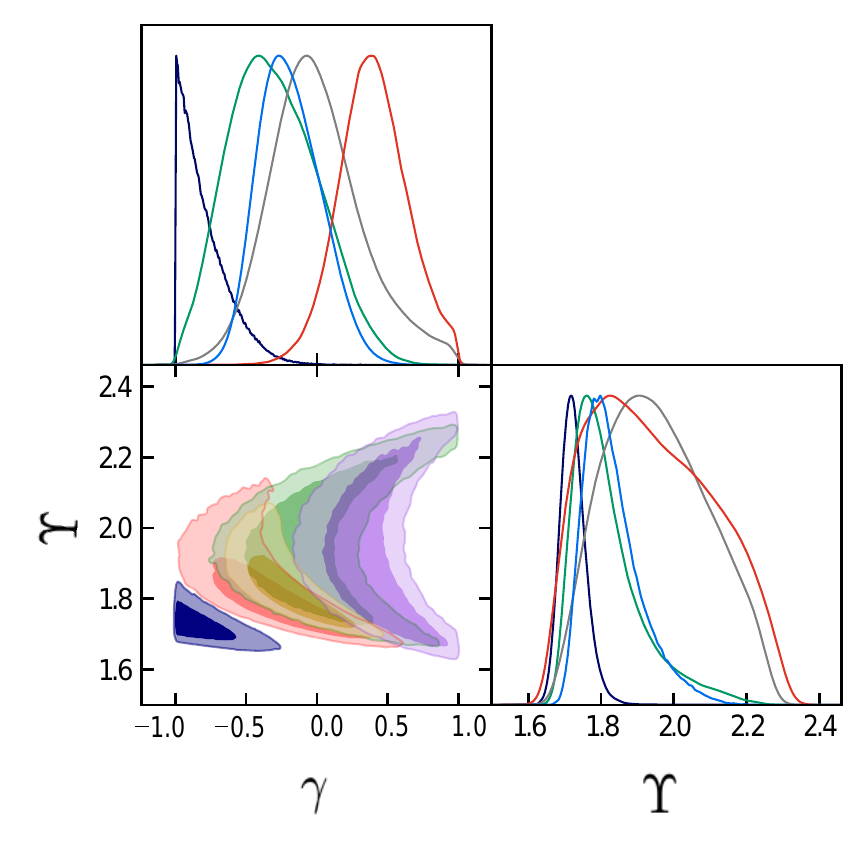}
\caption{Posterior probability distribution functions of $\gamma$ and $\Upsilon$ considering each sub-sample of strong gravitational lensing according to mass distribution. The blue outline corresponds to {\bf S1}, the red one to {\bf S2}, the green and purple ones to {\bf S3} and {\bf S4}, respectively,  and the yellow one to {\bf S5}.}
\label{dlriess}
\end{figure}

\begin{figure}[htb!]
\centering	
\includegraphics[scale=0.9]{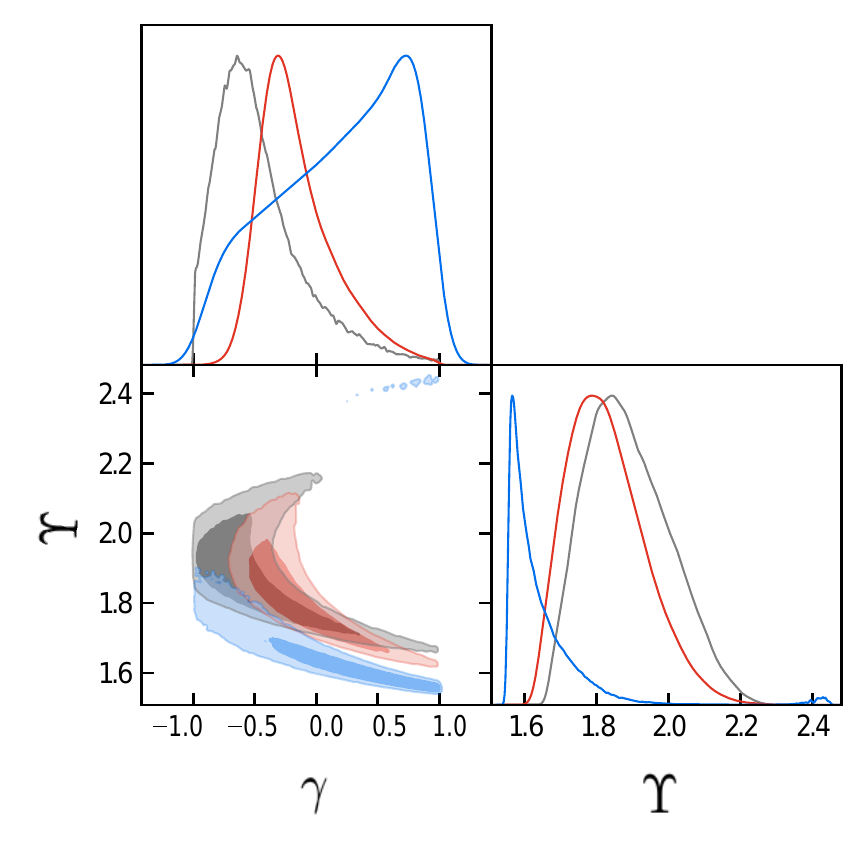}
\caption{Posterior probability distribution functions of $\gamma$ and $\Upsilon$ considering each sub-sample of strong gravitational lensing according to lens redshift. The grey outline corresponds to {\bf S6}, the blue one to {\bf S7}, and the red one {\bf S8}. }
\label{dlriess}
\end{figure}

\begin{table*}
\centering
	\begin{tabular}{|c|c|c|c|c|c|c|c|} \hline
Sub-Sample & $N$ & $\gamma$ & $\Upsilon$ & $\chi_{\mathrm{red}}^{2}$  & $\gamma'$ & $\Upsilon'$ & ${\chi'}_{\mathrm{red}}^{2}$ \\  \hline

{\bf S1} & $28$ & $-0.92 \pm 0.085$ & $1.73 \pm 0.020$ & $1.938$ & $-0.83 \pm 0.165$ & $1.72 \pm 0.035 $ & $0.781$ \\ \hline
{\bf S2} & $40$ & $-0.19 \pm 0.190$ & $1.77 \pm 0.035$ & $1.524$ & $-0.33 \pm 0.325$ & $1.80 \pm 0.095$ & $0.337$ \\ \hline
{\bf S3} & $18$ &$-0.13 \pm 0.155$ & $1.97 \pm 0.095$ & $1.638$ & $-0.02 \pm 0.315$ & $1.95 \pm 0.150$ & $0.351$  \\ \hline
{\bf S4} & $11$ & $+0.38 \pm 0.175$ & $1.85 \pm 0.125$ & $0.475$ & $+0.38 \pm 0.240$ & $1.93 \pm 0.180$ & $0.181$ \\ \hline
{\bf S5} & $72$ & $-0.16 \pm 0.140$ & $1.81 \pm 0.035$ & $1.759$ & $-0.21 \pm 0.220$ & $1.82 \pm 0.075$ & $0.386$\\ \hline
{\bf S6} & $55$ & $-0.69 \pm 0.120$ & $1.92 \pm 0.065$ & $2.203$ & $-0.53 \pm 0.340$ & $1.88 \pm 0.125$ & $0.418$  \\ \hline
{\bf S7} & $17$ & $+0.18 \pm 0.585$ & $1.63 \pm 0.075$ & $4.215$ &  $+0.25 \pm 0.630$ & $1.62 \pm 0.085$ & $2.229$ \\ \hline
{\bf S8} & $39$ & $-0.30 \pm 0.125$ & $1.88 \pm 0.075$ & $2.379$ & $-0.21 \pm 0.295$ & $1.81 \pm 0.105$ & $0.692$ \\ \hline
{\bf Full Sample} & $111$ &  $-0.44 \pm 0.105$ & $1.84 \pm 0.035$ & $2.664$ &  $-0.45 \pm 0.170$ & $1.81 \pm 0.050$  & $0.844$  \\
\hline    
	\end{tabular}
	\caption{Values of $\gamma$, $\Upsilon$ and $\chi_{\mathrm{red}}^{2}$ at $1\sigma$ of confidence level. The prime denotes the values of $\gamma$, $\Upsilon$ and $\chi_{\mathrm{red}}^{2}$ when the intrinsic error $\sigma_{int} \approx 12.22 \%$ is included in the analysis.}. 
\end{table*}

We note that considering different sub-samples leads to different results for both parameters $\gamma$ and $\Upsilon$. The sub-samples {\bf S1}, {\bf S4}, {\bf S6}, and {\bf S8} presented high values of $\gamma$ at $1\sigma$ of confidence level, suggesting a possible variation in $c$ and $\alpha$ with time. On the other hand, results from the analyses that only considered the sub-samples {\bf S2}, {\bf S3}, {\bf S5} and {\bf S7} are consistent within $1\sigma$ suggesting no variation in $\alpha$ or $c$. However, the sub-samples {\bf S4} and {\bf S7} yield positive values of $\gamma$ and also the lowest and highest value of $\chi_{\mathrm{red}}^{2}$ at $1\sigma$, respectively. On the other hand, the sub-samples {\bf S3} and {\bf S6} point to the highest values of $\Upsilon$, close to the limit $\Upsilon = 2$. Therefore, our analysis shows that the assumptions used for the lens mass distribution model are not accurate for the SGL data sets considered in this paper and therefore prevents us to obtain more conclusive results on the variation in $\alpha$ and $c$ with the method proposed in this paper.

\section{Conclusions}

According to the theory of relativity, the speed of light is the upper limit at which conventional matter, energy, or any signal carrying information can travel through space. Probing its invariance constitutes, therefore, a crucial test for observational cosmology. In this paper, we proposed a new method to investigate a possible time variation of $c$ assuming at the same time a possible time variation of the fine-structure constant, being both variations related. For this, we used strong gravitational lensing systems and type Ia of supernovae observations in a specific redshift range.

Our method relies in a general approach to describe the mass distribution of lens-type galaxies and the assumption of the runaway dilaton model to describe the variation in $\alpha$ with time. In this way, using MCMC methods, new limits on $\gamma$ and on $\Upsilon$ could be established when more accurate data are available.

In the present analysis, we split the full sample into five sub-samples according to the lens stellar velocity dispersion (low, intermediate, and high $\sigma_{ap}$) and into three sub-samples according to the lens redshift. The results pointed to a non-negligible scattering, however, they reinforced the need for segregating the lenses and analyzing them separately. Comparing the values of $c(z_l)$ obtained here with the current value measured on Earth, we conclude that {\bf it} is difficult to achieve competitive results with current astronomical observations located at different redshifts. Nonetheless, we stress that the main achievement of this paper is to propose a new method to measure the possible variation in $\alpha$ and $c$ with time with SGL and SNe Ia data. In a near future, more accurate datasets will be available to apply this method like the ones from the X-ray survey eROSITA \cite{eROSITA}, which is expected to detect 100.000 galaxy clusters approximately, along with follow-up optical and infrared data from the EUCLID mission, Nancy Grace Rowan space telescope, and Vera Rubin LSST that will detect a huge amount of strong gravitational lensing systems.

\section{Acknowledgments}
S.L is supported by grant PIP 11220200100729CO CONICET and grant 20020170100129BA UBACYT.  JEG acknowledges financial support from the Conselho Nacional de Desenvolvimento Cient\'{i}fico e Tecnol\'{o}gico CNPq (Grants no. 165468/2020-3).

\label{lastpage}
\end{document}